\documentclass{elsart}
 
\usepackage{amssymb,graphicx}

\newcommand{\etc}{$\ldots$}
\newcommand{\NIMR}[1]{Nucl. Instr. and Meth. {\bf #1}}
\newcommand{\dedx}{$\displaystyle{\d E/\d X}$} 
\newcommand{\Ln}[1]{\ln{\left(#1\right)}}
\newcommand{\elm}[2]{$^{#2}$#1}
\newcommand{\spc}{\quad}
\newcommand{\CsI}{CsI(Tl)}
\newcommand{\dpt}{\partial}
\newcommand{\dde}[1]{{\dpt\Delta E\over\dpt #1}}
\newcommand{\ddel}[1]{{\dpt\Delta E/\dpt #1}}
\newcommand{\dE}[1]{{\dpt E\over\dpt #1}}
\newcommand{\zal}{\lambda Z^{\alpha}A^{\beta}}
\newcommand{\zalm}{\left(\zal\right)^{\mu+\nu+1}}
\newcommand{\pgem}{(gE)^{\mu+\nu+1}}
\newcommand{\amuenu}{Z^2A^\mu (gE)^\nu}
\newcommand{\ovm}{\over \mu+\nu+1}
\newcommand{\Gm}{G^{{1\ovm}-1}}
\newcommand{\D}{\displaystyle}

\begin{document}
\begin{frontmatter}

\title{A new functional for charge and mass identification in 
     $\Delta$E-E telescopes}

\author{L. TASSAN-GOT}
\address{Institut de Physique Nucl\'{e}aire, 
    IN2P3-CNRS-Universit\'{e} Paris-Sud, 91406 Orsay, France}

\begin{abstract}
We propose a new functional for the charge and mass identification
in $\Delta E-E$ telescopes. This functional is based on the Bethe 
formula, allowing safe interpolation or extrapolation in regions 
with low statistics. When applied to telescopes involving detectors 
delivering a linear response, as silicon detectors or ionization 
chambers, a good mass and charge identification is achieved. For 
other detectors, as caesium-iodide used as a final member of a 
telescope, a good accuracy is also obtained except in the
low residual energy region. A good identification is however recovered
if a non-linear energy dependence of the light output is included.
\end{abstract}

\end{frontmatter}

\section{Introduction}

Stacks of detectors, called telescopes, measuring the energy loss and
residual energy of charged particles have been used for a long time to 
get charge and mass identification, and also energy, of such particles.  
Several combinations of detectors have been used for this purpose~: 
ionization chambers, silicon detectors, plastic scintillators, 
thallium-activated caesium-iodide scintillators (\CsI) read by
photomultipliers or photodiodes. The 
identification is obtained by plotting the energy loss in one or several
components of the detector stack versus the residual energy released in 
the detector in which the particle has stopped. In such a plot events of
a given charge and mass cluster around identification lines.
 Two methods can then be implemented to extract the identification for 
each event~:
\begin{itemize}
\item
 interactive drawing of lines on top of ridges corresponding to a given
 charge or mass, or contours around events clustered around these 
 ridges, any event being identified from its distance to ridge lines or
 its inclusion in one contour,
\item
  fit of the ridge lines with a functional in which $Z$ and $A$ enter as
  parameters, the identification being obtained by inversion of the
  functional for given $\Delta E$ and $E$ in order to extract $Z$ and 
  $A$.
\end{itemize}

Whereas the first method is probably more powerful and allows to face 
any situation, it suffers two main drawbacks~: it doesn't deliver any
extrapolation in regions of low statistics, and it is human-time 
consuming because each identification line has to be accurately drawn. 
This last aspect becomes really a concern when multidetectors are used, 
in which hundreds of such telescopes are involved.  The second method 
doesn't suffer these inconveniences provided that the functional 
accurately models the data. Identification functionals have already been
used in the past~\cite{Goulding}. Due to their limited range of 
application, some extensions have been added later~\cite{Butler}.
The purpose of this work is to propose a more complex functional based
on physical grounds in order to allow accurate modelling of the data and
safe extrapolations in regions where no data are present. 

The second section is devoted to the derivation of a functional based on
Bethe's formula with a power law velocity dependence, and to its main
properties. The third section proposes a phenomenological extension of
this functional for data departing from the simple restriction of 
Bethe's formula. In the fourth section we will show how the introduction
of a non-linear term for the amplitude of the light of a \CsI\ crystal 
allows a good charge identification up to $Z=50$ for telescopes 
involving such detectors.

\section{Basic functional}

  We assume that the stopping power of a fragment of energy 
${\mathcal E}$, mass $A$ and nuclear charge $Z$ in a detecting medium 
takes the simple form~:
\begin{equation}
{\d {{\mathcal E}}\over \d X} = {Z^2\over f({\mathcal E}/A)}
\label{eq:dedx}
\end{equation}
This formula can be straightly derived from Bethes's 
formula~\cite{Bethe} when the charge state inside the stopping medium
is strictly equal to the nuclear charge $Z$. Its range of validity is
restricted to light ions with energy sufficient to insure that they
are fully stripped. In particular the Bragg zone of the stopping power 
curve, where the mean charge state is no more constant, is not addressed 
by this formula.

We also define the integral $F$ of $f$ as~:
\begin{equation}
F(x) = \int\limits_0^x f(t)\,\d t\label{eq:int}
\end{equation}

By integration of equation~(\ref{eq:dedx}) one obtains the range-energy
relation~:
\begin{equation}
 F({\mathcal E}/A) = {Z^2\over A}X \label{eq:range}
\end{equation}
This last relation can be applied to the case of telescopes made of a
first detector of thickness $\Delta X$. The incoming fragment releases 
a part $\Delta E$ of its energy in this detector, and its residual 
energy $E$ in the rear detector for a residual range $X$.

When applied to the total energy and to the residual energy, 
relation~(\ref{eq:range}) reads~:
\begin{eqnarray*}
  \left\{
 \begin{array}{rcl}
  \D{Z^2\over A}(X+\Delta X) &=& F\left(\D{E+\Delta E\over A}\right)\\
  \D{Z^2\over A}X &=& F\left(\D{E\over A}\right)
 \end{array}
 \right.
\end{eqnarray*}
which, by elimination of $X$, delivers the $E$-$\Delta E$ relation~:
\begin{equation}
\Delta E = A\left\{F^{-1}\left[F(E/A)+{Z^2\over A}\Delta X\right]-
  {E\over A}\right\}\label{eq:ede0}
\end{equation}
This formula is rigorous as long as the stopping power takes the
form~(\ref{eq:dedx}). However to reach a practical use we need
to specify the function $f$. We will choose a form close to Bethe's 
behaviour but simpler in order to obtain and analytical form of the 
integral inverse $F^{-1}$.

\subsection{Specialisation of the function $f$}

In the case of Bethe's formula in the non-relativistic domain, 
$f({\mathcal E}/A)$ is almost proportional to ${\mathcal E}/A$ because 
the logarithmic term can be considered as constant. Therefore we will 
choose a power law dependence~:

\begin{equation}
 f({\mathcal E}/A) = ({\mathcal E}/A)^{\mu}\label{eq:spb}
\end{equation}
with $\mu\approx 1$.
This particular form of $f$ is advantageous because it leads to an
analytical form of~(\ref{eq:ede0}) which reduces to~:
\begin{equation}
\Delta E = \left[E^{\mu+1}+(\mu+1)Z^2A^{\mu}\,\Delta X\right]
^{1\over \mu+1}-E\label{eq:ede1}
\end{equation}
One can notice that this relation is strictly equivalent to the 
functional used in~\cite{Goulding} because both are based on the same 
hypotheses.

\subsection{Properties of the $E$-$\Delta E$ relation}

By taking the derivative of expression~(\ref{eq:ede1}) versus $E$ at
$E=0$ one gets the slope at the starting point of the identification 
line which turns out to be equal to -1 independently of $Z$ and $A$. 
For signals delivered by real detectors this slope includes the ratio
of electronic gains. However, in case of detectors exhibiting a linear 
response as silicon detectors, this constancy of the starting slope is 
generally achieved. Conversely the verification of this property is a 
good indication of the validity of approximations brought 
by~(\ref{eq:dedx}).

By setting $E=0$ in~(\ref{eq:ede1}) one obtains the series of crossing 
points of identification lines with the energy loss axis~:
\begin{equation}
\Delta E_0 = \left[(\mu+1)\,\Delta X\right]^{1\over \mu+1}
  Z^{2\over \mu+1}A^{\mu\over \mu+1}\label{eq:de0}
\end{equation}
In particular for $\mu=1$, close to reality, these starting energy 
losses are proportional to~$Z\sqrt{A}$.

At high energy  the second term between $[\ldots]$ in 
equation~(\ref{eq:ede1}) becomes small, compared to the first one, 
and an expansion to first order can be performed~:
\begin{equation}
\Delta E_{\infty} = \Delta X{Z^2\over (E/A)^\mu}\label{eq:dei}
\end{equation}
which is proportional to the stopping power, as expected.

\subsection{Power law functional}

If we assume that the detector response is linear, and that data comply
to relation~(\ref{eq:ede1}), then they should be fitted by the 
following function~:
\begin{equation}
\Delta E = \left[(gE)^{\mu+1}+\left(\lambda\, Z^{2\over \mu+1}
     A^{\mu\over \mu+1}\right)^{\mu+1}\right]^{1\over \mu+1}-gE
     \label{eq:fit0}
\end{equation}
where the parameters are $g$ (ratio of electronic gains),
$\lambda$ which includes the thickness of the first detector and $\mu$
which is close to 1. Furthermore $E$ and $\Delta E$ are net signals 
from which pedestals have been subtracted. We then obtain a 3-parameter
formula, or a 5-parameter formula if we also fit on the $\Delta E$ and 
$E$ pedestals.

To guarantee a convergence of the fit in all situations one should 
provide reasonable starting values and impose constraints on the range 
of the parameters. In particular $\mu$ should lie between 0.5 and 1.5 
with a starting value $\mu_0=1$, the initial value $\lambda_0$ of 
$\lambda$ should be determined from the starting point of an 
identification line, and $g_0$ should be obtained from the slope 
at the same point. The reader is referred to the appendix for  
technicalities related to the fit procedure.

 Once parameters have been determined by the fitting procedure, the
function~(\ref{eq:fit0}) can be used to extract $Z$ and eventually $A$
associated to any $\Delta E-E$ pair. This is achieved by an analytical
inversion of equation~(\ref{eq:fit0}) which delivers the quantity 
$Z^2A^{\mu}$ as early observed~\cite{Goulding}. If one is only 
interested in the charge identification $A$ is set dependent on $Z$, 
the simplest prescription being $A=2Z$, and equation~(\ref{eq:fit0}) 
is solved for $Z$ and the solution is projected onto the nearest 
integer number $Z_i$. Furthermore if the mass identification is 
required, the value $Z_i$ is taken for $Z$ and the equation is solved 
for $A$.

\begin{figure}
\begin{center}
\includegraphics{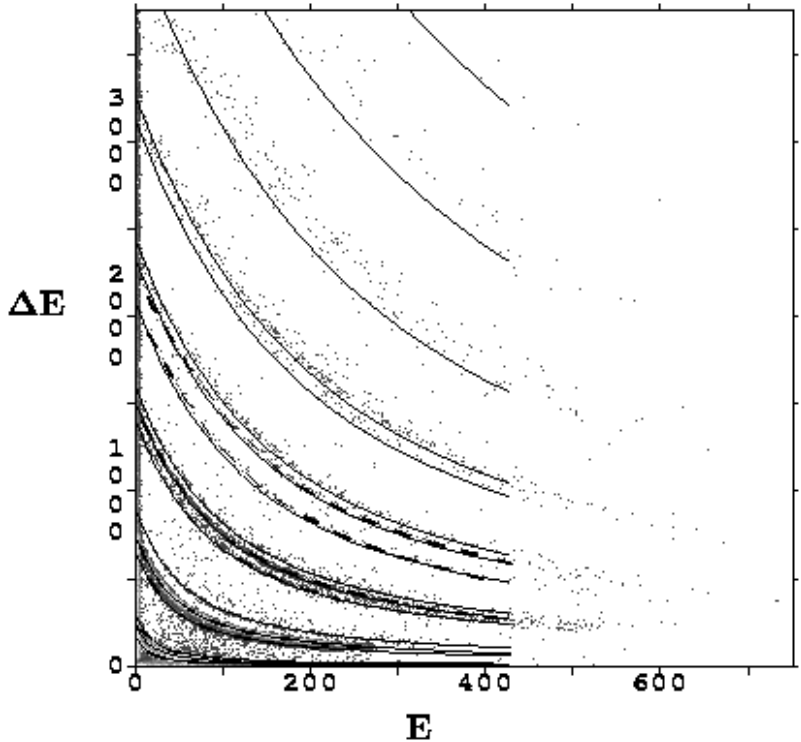}
\end{center}
\caption{\label{fig:sisiexp}
$\Delta E-E$ map from a silicon-silicon  telescope.
The 8~dashed lines, from hydrogen to beryllium, are the reference lines 
used for the 5-parameter fit of equation~(\ref{eq:fit0}), which in turn
generates the thin continuous lines. For $Z=6,7,8$, $A=2Z$ is assumed.}
\end{figure}

\begin{figure}
\begin{center}
\includegraphics{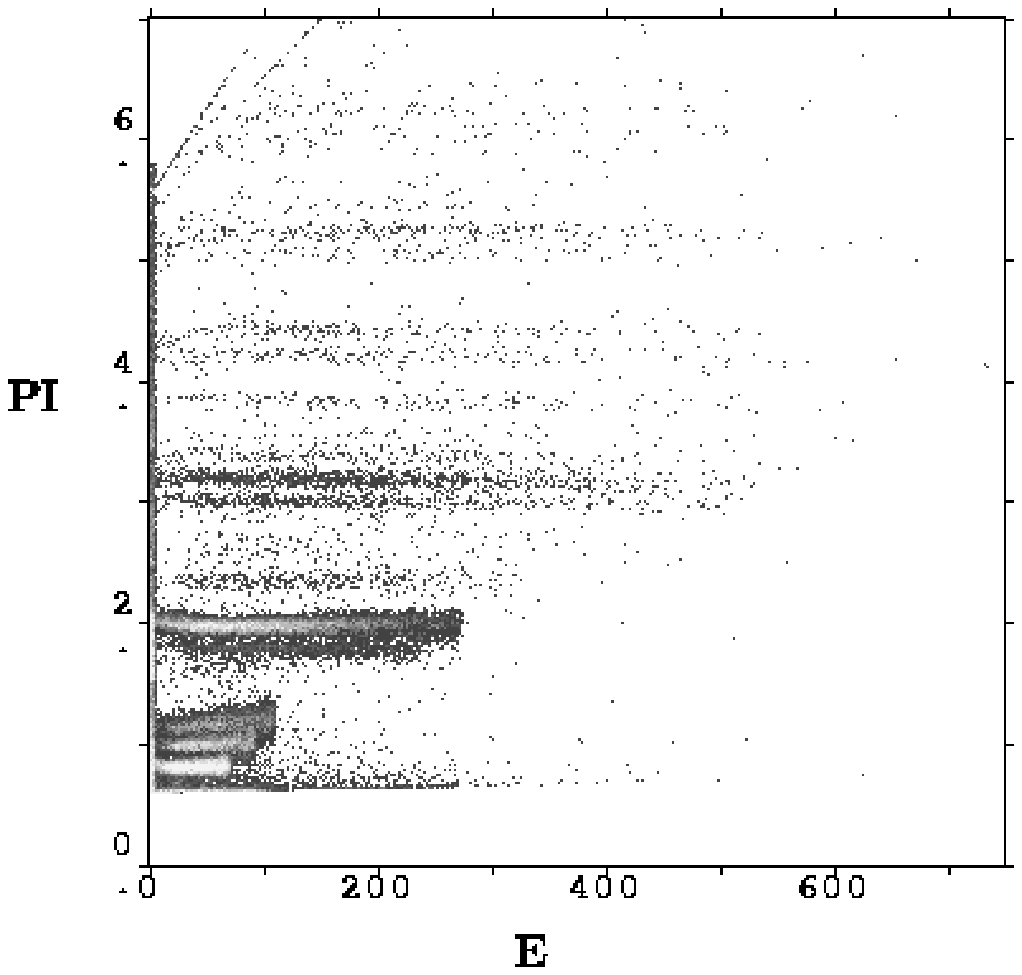}
\end{center}
\caption{\label{fig:sisifi1}
Plot of the particle identification variable defined as
$PI=Z_i+0.2(A-2Z_i)$ where $Z_i$ is the nearest integer deduced from 
the functional~(\ref{eq:fit0}) setting $A=2Z$, and $A$ is the mass 
which solves~(\ref{eq:fit0}) when $Z=Z_i$. 
}
\end{figure}

 As an example figure~\ref{fig:sisiexp} shows the $\Delta E-E$ map 
obtained with a calibration module of INDRA, made of a silicon detector
of thickness 70~$\mu$m and a lithium-diffused silicon detector 2~mm 
thick, placed in front of a \CsI\ crystal used as a veto removing all 
particles punched through the silicon-lithium detector. The 8~dashed 
lines are the reference lines, from hydrogen to beryllium, used as 
references for the 5-parameter fit. The result of the fit is shown on 
figure~\ref{fig:sisifi1} which displays the particle identification 
$PI=Z_i+0.2(A-2Z_i)$ versus the energy released in the silicon-lithium 
detector. In can be verified that distorsions are very small and that 
the extrapolation toward boron an carbon elements is correct, 
indicating that the power law formula models correctly the data in 
this range.

\begin{figure}
\begin{center}
\includegraphics{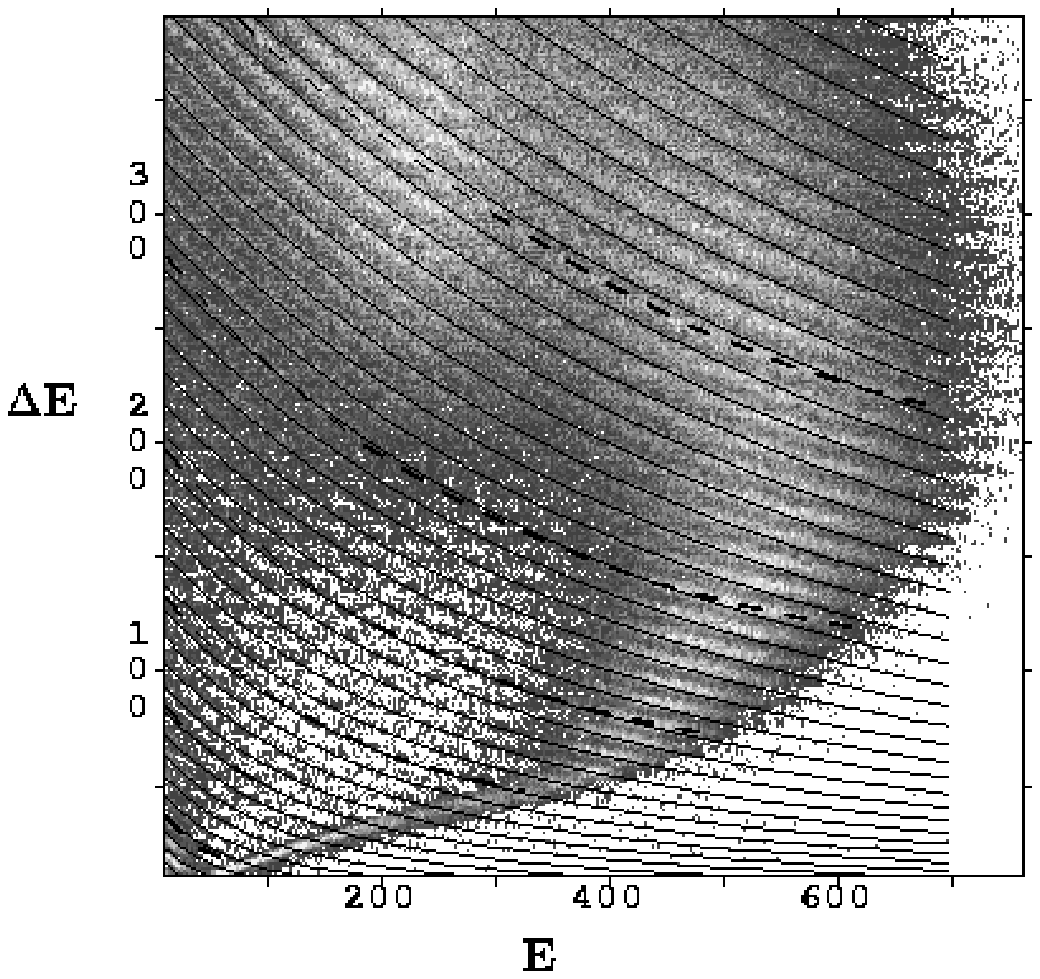}
\end{center}
\caption{\label{fig:sisigan1}
$\Delta E-E$ map obtained with a 150~$\mu$m-750~$\mu$m
silicon telescope for the \elm{Pb}{208}+\elm{Si}{nat} reaction at
29~MeV/n~\cite{Morjean}. More populated areas are indicated by clearer
zones. Thick dashed lines drawn for $Z=8,13,17,22,28,37$
are used as a reference for the 5-parameter fit with
relation~(\ref{eq:fit0}) which generates the thin lines.}
\end{figure}

However it has already been noticed in the past~\cite{Butler,Shimoda}
that  for a wider $Z$ range, the above formula must be extended.
The stopping power may depart from the behaviour ruled by
relations~(\ref{eq:dedx}) and~(\ref{eq:spb}), moreover the 
response of some detectors is not completely linear~: pulse height 
defect in silicon detectors, light response of \CsI\  detectors\etc\ 
Figure~\ref{fig:sisigan1} shows the $\Delta E-E$ map obtained with 
a telescope made of 2~silicon detectors, of thicknesses equal to
150~$\mu$m and 750~$\mu$m,  in the case of a \elm{Pb}{208} beam at 
29~MeV/n~\cite{Morjean}.
Thick dashed lines drawn on top of ridges associated to
$Z=8,13,17,22,28,37$ are used as a reference for the 5-parameter fit 
with relation~(\ref{eq:fit0}) which generates the thin lines. It turns
out that the functional hardly reproduces the data in the whole range~: 
not only its $Z$ dependence is approximative, but also the shape of 
the energy dependence for a given line.

Therefore an enrichment of~(\ref{eq:fit0}) is highly desired in order
to benefit from more degrees of freedom. The extension proposed 
hereafter is not based on theoretical considerations but rather on 
phenomenological requirements.

\section{Extended functional}

One can see from equation~(\ref{eq:fit0}) that $\mu$ rules the ordinates
of starting points of identification lines. Once $\mu$ has been fitted 
to reproduce these data, the rapidity of the transition from the low 
energy domain dominated 
by~$\displaystyle\lambda\, Z^{2\over \mu+1} A^{\mu\over \mu+1}$
to the high energy region, proportional to the stopping power, is fixed.
This could lead to a too strong constraint when applied to real data.
One could introduce an additional parameter for the $Z$ exponent but 
this would still link the $Z$ dependencies in the low and high energy 
regions. As regards the $A$ exponent, it can't be touched because it 
insures the right asymptotic behaviour at high energy, toward the 
stopping power. 

The ideal solution should meet the following prescriptions~: 
independent variations in $Z$ and $A$ in the low and high energy 
regimes, possibility of tuning separately the rapidity of the transition
between these regimes, correct asymptotic behaviour at high energy. 
This is achieved by the addition of new phenomenomogical term.

We define the extended functional as~:
\begin{equation}
\Delta E = \left[(gE)^{\mu+\nu+1}+\left(\lambda Z^{\alpha}
     A^{\beta}\right)^{\mu+\nu+1}+\xi Z^2A^\mu 
     (gE)^\nu\right]^{1\over \mu+\nu+1}-gE\label{eq:fit1}
\end{equation}
which reduces to the basic functional~(\ref{eq:fit0}) when one
substitutes $\mu$ for $\mu+\nu$ and sets $\alpha=2/(\mu+1)$,
$\beta=\mu/(\mu+1)$ and $\xi=0$, or even more simply if one
sets~: $\nu=0$ and $\lambda=0$.

The $g$ parameter still represents the slope of identification lines 
at their starting points (at $E=0$), equal for all lines as for the 
power law formula. The energy loss value at the starting points writes~:
$\Delta E_0 = \lambda Z^{\alpha}A^{\beta}$, showing that this series of
points essentially determines the $\lambda$, $\alpha$ and $\beta$ 
parameters.

At higher energies an expansion of~(\ref{eq:fit1}) to first order can 
be carried out, leading to~:
\begin{displaymath}
\Delta E_{\infty} = {1\over\mu+\nu+1}\times{1\over (gE/A)^\mu}\times
 \left[\xi Z^2+{(\lambda Z^{\alpha})^{\mu+\nu+1}A^{\beta(\mu+\nu+1)-
  (\mu+\nu)}\over (gE/A)^\nu}\right]
\end{displaymath}
One can notice that at high energy the \dedx\ is recovered if
$\mu$ stays in the vicinity of 1, and that $\xi$ governs the amount of
energy loss in this region.  Moreover one benefits from the second term
inside the square brackets which allows, through the $\nu$ parameter, 
to change the shape of curves in the region of intermediate energy.

The extended formula is a 7-parameter functional 
($\lambda$, $\alpha$, $\beta$, $\mu$, $\nu$, $\xi$ and $g$) or a 
9-parameter one if both pedestals are also considered as parameters of 
the fit. The convergency of the fit is always insured only if good 
starting values are supplied and if constraints are applied to the range
of parameters. More precisely one should adopt~:
$\alpha_0=\mu_0=1$, $\beta_0=0.5$, $\nu_0=1$, $\lambda_0$ determined 
from the starting point of an identification line, $g_0$ from its 
starting slope, and $\xi_0$ from one point taken at high energy. The 
ranges should be restricted to~:
\spc $0.5<\alpha<1.5$,\spc $0.2<\beta<1$,\spc $0.2<\mu<1.5$,
\spc $0.1<\nu<4$, \spc $\lambda_0/4<\lambda<4\lambda_0$,
\spc $g_0/4<g<4g_0$,\spc $0<\xi$.
In the case where no mass identification is given, the number of 
parameters is decremented by one~: $\beta$ is kept equal to 0.5 and 
$A$ is taken as a function of $Z$, the simplest relationship being 
$A=2Z$.

 Unlike the power law formula, the extended formula can't be 
analytically  inverted to extract $Z$ and $A$ from a $\Delta E-E$ pair.
This can only be done numerically and the fastest algorithm is probably 
the Newton-Raphson method which however needs the derivatives 
of~(\ref{eq:fit1}) with respect to $Z$ and $A$.

\begin{figure}
\begin{center}
\includegraphics{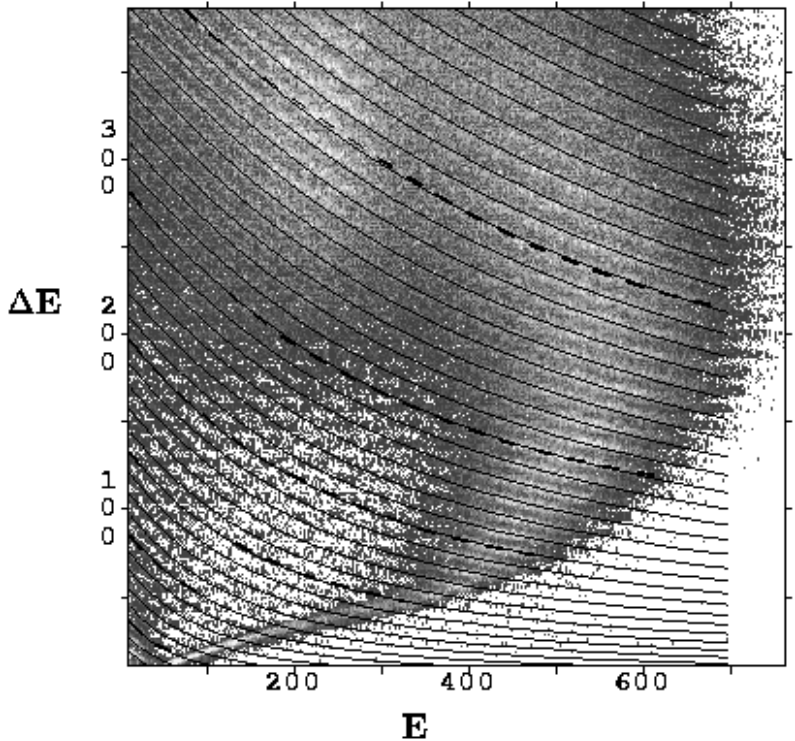}
\end{center}
\caption{\label{fig:sisigan2}
Same as figure~\ref{fig:sisigan1} but thin
lines result now from the fit based on the 8-parameter extended 
functional~(\ref{eq:fit1}), assuming $A=2Z$.}
\end{figure}

\begin{figure}
\begin{center}
\includegraphics{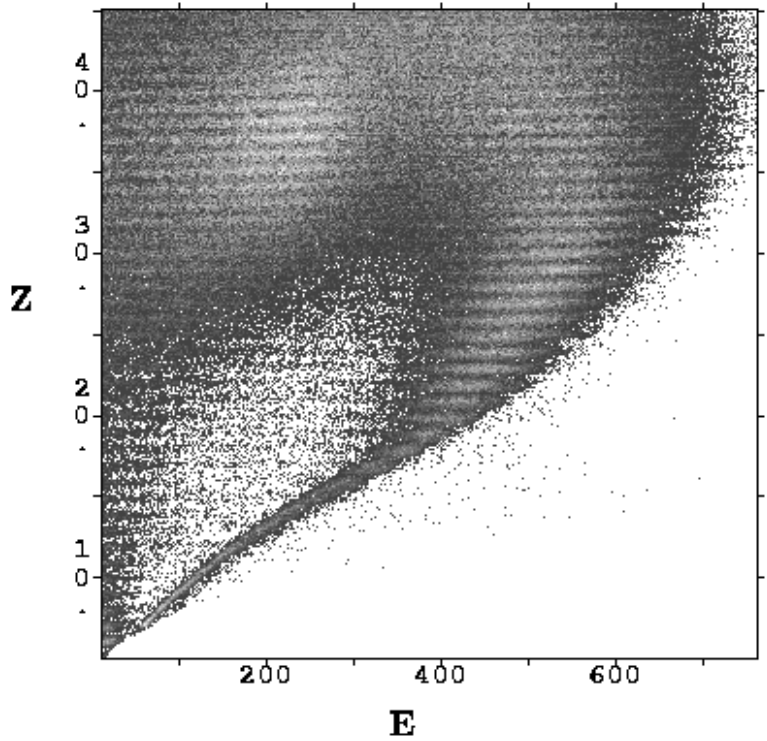}
\end{center}
\caption{\label{fig:sisigan3}
For the same data as in figures~\ref{fig:sisigan1}
and~\ref{fig:sisigan2}, the $Z$ extracted from relation~(\ref{eq:fit1})
is plotted versus the residual energy to exhibit possible distorsions.}
\end{figure}

 The quality of a fit based on this extended formula with 8~parameters,
assuming $A=2Z$ and $\beta$ arbitrarily set, is shown on
figure~\ref{fig:sisigan2} where it appears that the shape of the curves
is nicely reproduced, and also the interpolation in regions where no 
reference line guides the fit procedure. Even the extrapolation toward 
higher $Z$'s is satisfactory, having in mind that in this region the 
simple hypothesis $A=2Z$ is unlikely to hold. Figure~\ref{fig:sisigan3} 
shows that only small distorsions remain when the identified $Z$ for 
each event is plotted versus the residual energy.

\begin{figure*}
\begin{center}
\includegraphics[width=13.5cm]{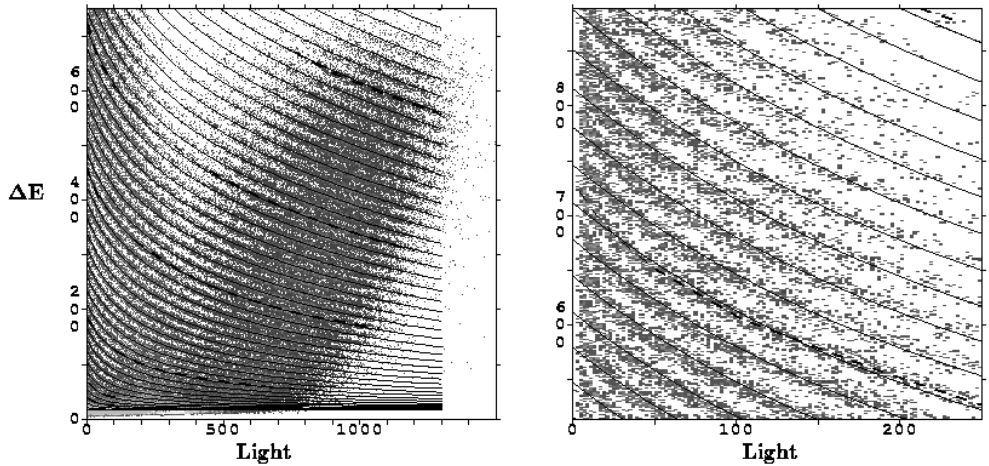}
\end{center}
\caption{\label{fig:lusipx}
$\Delta E-h$ plot for a silicon-\CsI\ telescope, the 
abscissa is the total light output reconstructed from the fast and slow
components of the \CsI. The thick dashed lines follow the ridges of 
charges $Z=4,6,12,22,30,40$ and the thin lines  result from the 
8-parameter fit based on the extended functional~(\ref{eq:fit1}) with 
$A=2Z$.  The right part is a zoom applied to the low light region of the
left map, including the line $Z=30$. The non-linearity of the light 
response shows up by the high curvature of data, not predicted by the 
functional.}
\end{figure*}

 We can also check the ability of this functional to reproduce the data
for telescopes using a \CsI\ crystal as a stopping detector, for which
the light response is not linear with the deposited energy and depends
on the nature of the particle. For this purpose we worked on the data
collected by a module of the second ring of the multidetector INDRA,
made of a 300~$\mu$m silicon planar detector followed by a \CsI\ of
depth equal to 14~cm~\cite{INDRA}. As the \CsI\  signal was only
partially integrated on 2~time intervals for the delivering of the
so-called slow and fast components, we didn't have at disposal a true
measurement of the total light signal, integrated over the duration of
the pulse. We reconstructed this total light signal $h$ by a combination
of slow and fast components following a procedure described 
in~\cite{Parlog}. 
Figure~\ref{fig:lusipx} is an illustration of a $\Delta E-h$ plot
collected with a tin beam. The thick dashed lines are drawn on top of
charges $Z=4,6,12,22,30,40$ and the thin lines  result from the 
8-parameter fit with $A=2Z$.  Despite the non-linearity of the
light response and its dependence with $Z$ the fit is in good agreement 
with the data, showing that the parameters of the functional are able to
partially compensate for these effects. However by looking carefully at
the low residual energy region, as displayed in the right part of
figure~\ref{fig:lusipx}, one can see that the curvature of data is much
more pronounced than predicted by the functional. This effect is due to 
the highly non-linear light response  which is almost quadratic in this 
region. Any improvemevent of the accuracy of generated identification
 lines in this region needs an explicit inclusion of such a 
 non-linearity.

\section{Light response of caesium-iodide crystal}

The idea for a new extension of functionals~(\ref{eq:fit0}) and
(\ref{eq:fit1})
is to substitute for $E$ its evaluation from the light output, in order 
to derive direct dependencies of the energy loss with the light output, 
the most part of non-linearities being carried by the relationship 
between light and energy.

We will take for the light response of \CsI\ crystals a dependence
deduced from the Birks formula~\cite{Birks}. If one denotes
$h$ the net light output after subtraction of the pedestal, its 
dependence with the energy deposition takes the form~\cite{Birks2}~:
\begin{equation}
h=E-\rho\Ln{1+{E\over\rho}}\qquad\hbox{where}\qquad \rho=\eta Z^2A
 \label{eq:Eh}
\end{equation}
and $\eta$ is a new parameter. No multiplicative constant is needed 
because the scaling factor is already included in the $g$ parameter
of the functional, this is equivalent to consider the energy and the 
light output as measured with the same unit. This formula is not the 
most accurate~\cite{Parlog} and particularly it discards the loss of 
quenching due to the delta electrons. We take it merely as a reasonable
way of accounting for the non-linearities
at low energy and high $Z$, hoping that the other parameters of the
functional will be able to compensate its partial inadequacy.

However one difficulty arises from the fact that equation~(\ref{eq:Eh})
cannot be analytically inverted to express $E$ as a function of $h$, as
required for its insertion into the functionals~(\ref{eq:fit0}) or 
(\ref{eq:fit1}). This is the reason why we adopted a slightly different 
formula for the inverse relation, imposing the same asymptotic 
behaviours at low and high energy. From~(\ref{eq:Eh}) we derive for 
the low and high energy regimes~:
\begin{eqnarray*}
 E\longrightarrow 0 & \qquad\Longrightarrow\qquad & 
  h\longrightarrow{E^2\over 2\rho}  \\
 E\longrightarrow\infty & \qquad\Longrightarrow\qquad & 
  h\longrightarrow E-   \rho\Ln{E\over\rho} 
\end{eqnarray*}
which exhibits the quadratic dependence at low energy.
By inverting these relations one gets~:
\begin{eqnarray*}
 h\longrightarrow 0 & \qquad\Longrightarrow\qquad & 
  E\longrightarrow\sqrt{2\rho h} \\
 h\longrightarrow\infty & \qquad\Longrightarrow\qquad & 
 E\longrightarrow h+\rho\Ln{h\over\rho}
\end{eqnarray*}
Now we seek for a formula which complies with the above limiting 
behaviours, and we can check that~:
\begin{equation}
E=\sqrt{h^2 +2\rho h\left[1+\Ln{1+{h\over\rho}}\right]}\label{eq:hE}
\end{equation}
fulfills this condition. Obviously relation~(\ref{eq:hE}) is not the 
strict inverse of (\ref{eq:Eh}), but it departs from it by 7.4~\% only
at maximum, and both become identical at low and high energy.

If we use this relation in combination with~(\ref{eq:fit1}), we obtain
now a functional with 8~parameters ($\lambda$, $\alpha$, $\beta$,
$\mu$, $\nu$, $\xi$, $g$ and $\eta$), and 2 eventually additional
parameters if pedestals in $h$ and $\Delta E$ are also adjusted.

\begin{figure*}
\begin{center}
\includegraphics[width=13.5cm]{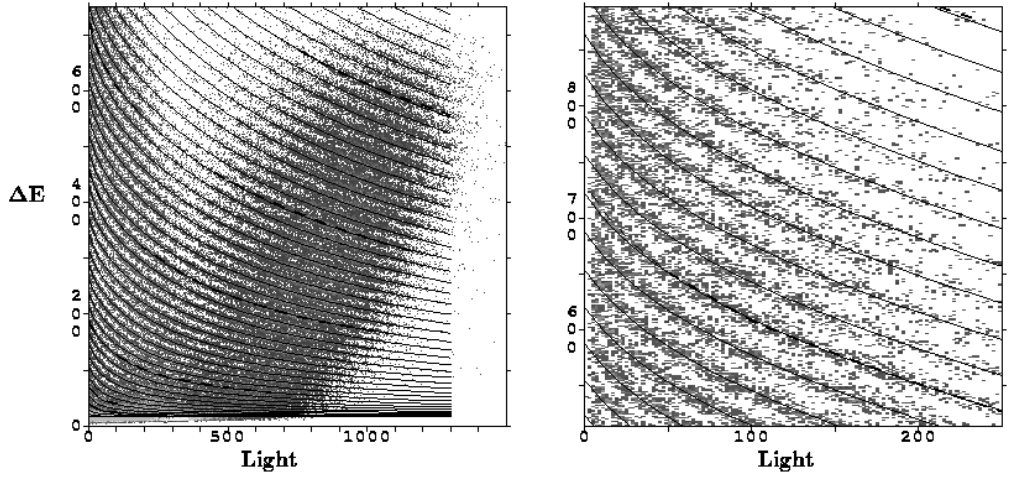}
\end{center}
\caption{\label{fig:lusipxx}
Same as figure~\ref{fig:lusipx} but the thin lines result
now from a 9-parameter fit combining relations~(\ref{eq:fit1}) and
(\ref{eq:hE}). The
curvature of data at low energy is now reproduced.}
\end{figure*}

Figure~\ref{fig:lusipxx} displays the calculated lines resulting from a
9-parameter fit in which $\beta$ has been kept constant as the charge 
identification only is given, and $A=2Z$ is assumed. It can be seen that
a satisfactory agreement is obtained for the shape of the curves. 
Particularly one can notice that the high curvature of the lines in 
the vicinity of the energy loss axis is well reproduced by the 
introduction of the light response and its quadratic dependence at low 
energy. Furthermore a stable extrapolation is deduced for the charges 
higher than 40, the highest line used as a reference for the fit.

\begin{figure}
\begin{center}
\includegraphics{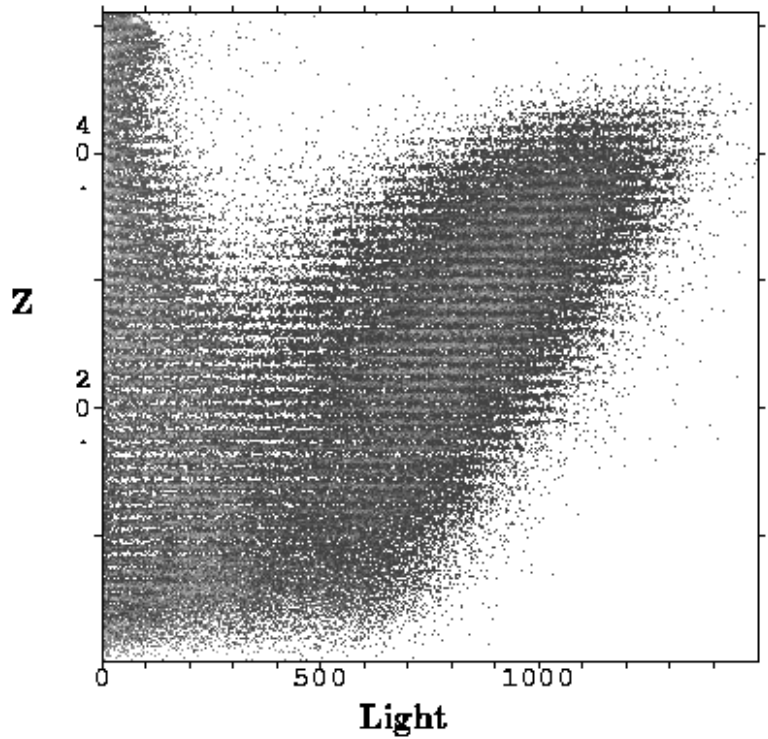}
\end{center}
\caption{\label{fig:lusipxxp}
For the same data as in figures~\ref{fig:lusipx}
and~\ref{fig:lusipxx}, the $Z$ extracted from relations~(\ref{eq:fit1})
and~(\ref{eq:hE}) is plotted versus the residual energy to exhibit 
possible distorsions.}
\end{figure}

 The quality of the charge identification can be checked on
figure~\ref{fig:lusipxxp} where the distorsions of the identified $Z$ 
versus the light output appear to be small even for a wide range of 
charges. The apparent loss of resolution at low $Z$'s simply reflects 
the coding granularity, the channel size in energy loss becoming too 
large for these low charges.

\section{Conclusion}

Starting from the usual power law formula we propose an extended
functional for the particle identification in $\Delta E-E$ telescopes,
allowing a good mass and charge identification with silicon detectors
on a wide range in energy and charge. Its domain of applicability 
remains however limited to situations where the Bragg region plays a 
minor role, in particular this excludes high $Z$'s detected in thin 
$\Delta E$ detectors. This functional can also be applied to specific 
cases where the $E$ signal is no longer linear with the energy 
deposition, as for example for silicon-\CsI\ telescopes in which the 
fast component of the \CsI\ is taken as a measure of the residual 
energy. This procedure has been  applied to the calibrating modules in 
the INDRA multidetector, however it appears to be accurate only on a 
range of a few charges. By using the total light output of \CsI, either
by integrating the whole light signal or by estimating it from the 
partial integrations associated to the slow and fast components, the
agreement between data and the extended formula is highly improved on a
large range of $Z$. Moreover the discrepancies which remain at low 
residual energy are removed by the explicit introduction of the 
non-linear light response of the \CsI\ crystal, and a good quality of 
the identification is obtained on a wide range in charge and energy. 
 This new functional opens the possibility of identifying particles in
silicon or silicon-\CsI\ telescopes by defining only a limited number of 
reference points, and allows to rely on the deduced interpolation and 
even on the extrapolation in low statistics regions. It should make 
easier the usual identification task of particle identification in the
case of charged particle multidetectors. The author expresses his 
aknowledgement to M.~Morjean and to the INDRA collaboration, for having 
provided their data before publication.

\begin{appendix}
\section{Appendix}
The fit algorithm relies on the local quadratic expansion of the
$\chi^2$. At each step a calculation is made for $\chi^2$,
its gradient, and an estimation of the hessian matrix obtained by
neglecting the second derivatives of the functional.  This 
approximation is powerful because it always delivers a definite
positive matrix which, in the vicinity of the minimum, becomes
very close to the true hessian matrix. With such a
prescription it is necessary to compute only the functional value
and its gradient at each data point. The minimization
proceeds iteratively and at each step the status of constraints
is checked.

 In order to insure a correct convergency 2~conditions have to be met~:
\begin{itemize}
\item
good starting values of the parameters have to be supplied as already
explained,
\item
for the minimization procedure the parameters have to be scaled 
in order to be of the order of unity.
\end{itemize}

For a practical use, the formulas giving the derivatives of the
functional~(\ref{eq:fit1}) against parameters are written hereafter.
By denoting~:
$$G = \left[\pgem+\zalm+\xi\,\amuenu\right]$$
the derivatives can be expressed as~:
\begin{eqnarray*}
  \dde{\lambda} &=& \Gm\zalm{1\over\lambda}\\
  \dde{\alpha} &=& \Gm\zalm\Ln{Z}\\
  \dde{\beta} &=& \Gm\zalm\Ln{A}\\
  \dde{g} &=& \left\{\Gm{1\over gE}\left[\pgem+
       {\nu\ovm}\xi\,\amuenu\right]-1\right\}\,E\\
  \dde{E} &=& \left\{\Gm{1\over gE}\left[\pgem+
       {\nu\ovm}\xi\,\amuenu\right]-1\right\}\,g\\
  \dde{\mu} &=& {\Gm\ovm}\left\{\pgem\Ln{gE}+
                 \zalm\Ln{\zal}+\vphantom{1^1\over 1}\right.\\
           &&\mbox{\hspace{2cm}}
           \left.\xi\,\amuenu\Ln{A}-{G\Ln{G}\ovm}\right\}\\
  \dde{\nu} &=& {\Gm\ovm}\left\{\pgem\Ln{gE}+\zalm\Ln{\zal}+
               \vphantom{1^1\over 1}\right.\\
           &&\mbox{\hspace{2cm}}
	   \left.\xi\,\amuenu\Ln{gE}-{G\Ln{G}\ovm}\right\}\\
  \dde{\xi} &=& {\Gm\ovm}\amuenu\\
  \dde{Z} &=& \Gm\left\{{\alpha\,\zalm+{2\ovm}
     \xi\,\amuenu \over Z}\right\}\\
  \dde{A} &=& \Gm\left\{{\beta\,\zalm+{\mu\ovm}
     \xi\,\amuenu \over A}\right\}
\end{eqnarray*}

The derivative $\ddel{E}$ is involved in the $E$-pedestal dependence.
It is also useful when $E$ is expressed from the light output
using~(\ref{eq:hE}).
The $\ddel{Z}$ and $\ddel{A}$ derivatives are ingredients of the
Newton-Raphson method for the determination of $Z$ and $A$
for a given $\Delta E-E$ pair whence the parameters have been determined.

Regarding the light response~(\ref{eq:hE}), the useful derivatives are~:
\begin{eqnarray*}
  \dE{h} &=& {h\over E}\left\{1+{1\over 1+h/\rho}+
        {1+\Ln{1+h/\rho}\over h/\rho}\right\}\\
  \dE{\rho} &=& {h\over E}\left\{{1\over 1+h/\rho}+\Ln{1+h/\rho}
   \right\}
\end{eqnarray*}

\end{appendix}

\end{document}